\documentclass[10pt]{iopart} 
\usepackage{hyperref}
\usepackage{color}
\usepackage[normalem]{ulem}
\usepackage{xcolor}
\usepackage{url}
\newif\ifheaders
\headerstrue

\newif\ifanswers
\answerstrue

\usepackage{graphicx}
\usepackage{xspace}

\newcommand{\acm}[1]{{\color{black} #1}}

\begin{document}
\title{Non-reversible Monte Carlo: an example of “true” self-repelling motion} \author{A.~C.~Maggs }

\ead{anthony.maggs@espci.fr} 


\address{CNRS UMR 7083, ESPCI Paris, Université PSL, 10 rue Vauquelin, 75005 Paris, France.} 

\begin{abstract} {We link the large-scale dynamics of non-reversible Monte Carlo
    algorithms as well as a lifted TASEP to an exactly soluble model of
    self-repelling motion. We present arguments for the connection between the
    problems and perform simulations, where we show that the empirical
    distribution functions generated from Monte Carlo are well described by the
    analytic solution of self-repelling motion.}
\end{abstract}
%
\ifheaders
\section*{Introduction}
\fi
Non-reversible Monte Carlo sampling as developed for hard disks~\cite{Wilson},
and its generalisation to arbitrary potentials~\cite{Manon, Harland_2017, Faulkner}
 has many wonderful properties which facilitate the equilibration of
large complex systems faster than conventional Monte Carlo, or molecular
dynamics algorithms~\cite{hexatic,hexatic2}. {Such Monte Carlo methods
do not implement the  historic choice of detailed balance, rather the
weaker condition of global balance. This leads to much greater freedom of choice for the implementer of algorithms.}
However, analytic understanding of the large-scale,
effective dynamics is lacking. This is in strong contrast to, for instance,
molecular dynamics which generate the Navier-Stokes equations at large
scales in a fluid. Extensive theoretical analysis of the mode structure of these
hydrodynamic equations gives us the slow hydrodynamic modes of
sound, vorticity and heat propagation, which limit the large-scale sampling of a
fluid.

What are the equivalent statements for event-chain Monte Carlo?  What are the
slow, hydrodynamic modes? This letter aims to present the continuum,
coarse-grained equations describing event-chain simulation and demonstrate their
equivalence to an exactly soluble model of a growing polymer.  Clearly, if one
understands the temporal evolution of the coarse-grained equations, it will be
possible to make exact statements on the evolution of densities and
correlations, and perhaps find better implementations in the future. The present
letter makes a direct link between event-chain Monte Carlo simulation and ``true'' self-repelling motion,~\cite{Amit} which
has been much studied using a variety of physical methods including
scaling~\cite{Bernasconi, Pietronero} and renormalisation~\cite{Obukhov_1983}.
Recently this model has been studied using advanced methods based on interacting
Brownian paths, which have led to exact, analytic results for the dynamical
properties~\cite{Wendelin, Toth, DUMAZ20131454}.

In this letter, we introduce the self-avoiding model, together with the exact
results for its time evolution. We then summarise the behaviour of a class of
non-reversible algorithms before arguing that there is a link
between these two dynamical systems. Finally, we present numerical evidence as
to the identity of distribution functions with data coming from event-chain
simulation of harmonic chains, as well as simulation of a {recently introduced} lifted TASEP (Totally
Asymmetric Simple Exclusion Process).

\ifheaders
\section*{True self-avoiding motion}
\fi
The ``true'' self-avoiding walk was introduced,~\cite{Amit} as a
dynamic model of polymer growth, in contrast to an equilibrated statistical
ensemble of equilibrated polymers. On a lattice, monomers are added successively
to a chain, trying to avoid places where the polymer has already passed. The
probability of  choosing a site $i$, which has been visited $L_i(t)$ times is
then
\begin{equation}
  p_i(t+1) = \frac{e^{- \lambda L_i(t) }} {\sum_{j} e^{- \lambda L_j(t)} }  \label{eq:disc}
\end{equation}
Where the sum is over all neighbours $j$ of the current position at time
$t$. {$\lambda>0$ measures the strength of the repelling interaction.}
The model has infinite memory, required to calculate the placement probabilities
of the new monomer. {The large-scale behaviour of the process is believed to be
independent of $\lambda$ for finite values of the parameter.}

It was argued,~\cite{Amit} that this dynamic process has a continuum limit so
that the effective, large-scale behaviour is of the form
\begin{eqnarray}
  \frac{d {\textbf X(t)} } {d t } &=& -\nabla L(t, {\textbf X}(t)) + \xi(t) \label{eq:motion} \\
  \frac {dL(t,{\textbf x}) }{dt} &=&\,\,  \delta ( {\textbf x} - {\textbf X(t)})  \label{eq:density}
\end{eqnarray}
The function, $L(t,{\textbf x})$, a \textit {local time}, cumulates memory as to
the occupation of the position ${\textbf X}(t)$.  {It is the
continuum analogue of  the discrete variable $L_i$ of eq.~(\ref{eq:disc}). We use
the symbol $L$ for both the discrete and continuum variables}.  {$\xi(t)$ in the
 formulation~\cite{Amit} corresponds to Gaussian noise, {though the
   mathematical literature shows that its exact form and existence is rather subtle}.} The growth of the end of the
growing polymer is then repelled by regions where $L(t,{\textbf x})$ has become
large, in a manner that is analogous to the original lattice model.  This
continuum model is amenable to many of the formal methods of field
theory~\cite{PELITI1984225,Obukhov_1983}.
 \acm{The model is also widely
  studied in the biophysical literature \cite{prx, biophy} where it is argued
  that the enhanced spreading due to self-interactions leads to improved spatial
  exploration of biological random walks by living organisms. In this context, it can be taken as
  a minimum model of self-interacting chemotaxis.} 

\begin{figure}
  \includegraphics[width=.65\columnwidth]{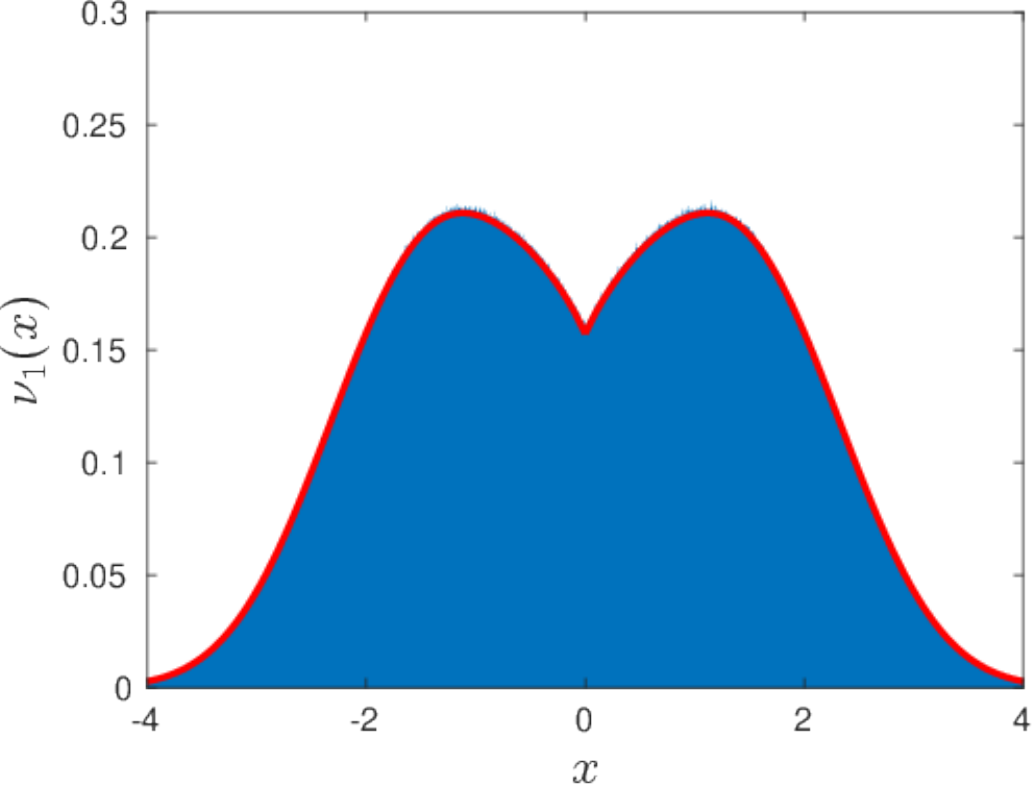}
  \caption{Scaling function for distribution, $\nu_1(x)$, eq.~(\ref{eq:nu1})
    curve in red, compared to simulation data where we have saved the index of
    the final point of a trajectory, blue histogram. The distribution is a
    function of only $|x|$, and is singular at $x=0$. The empirical distribution has a
    standard deviation  of $\sim16,500$ sites.  
}\label{fig:nu1}
\end{figure}

\begin{figure}
  \includegraphics[width=.65\columnwidth]{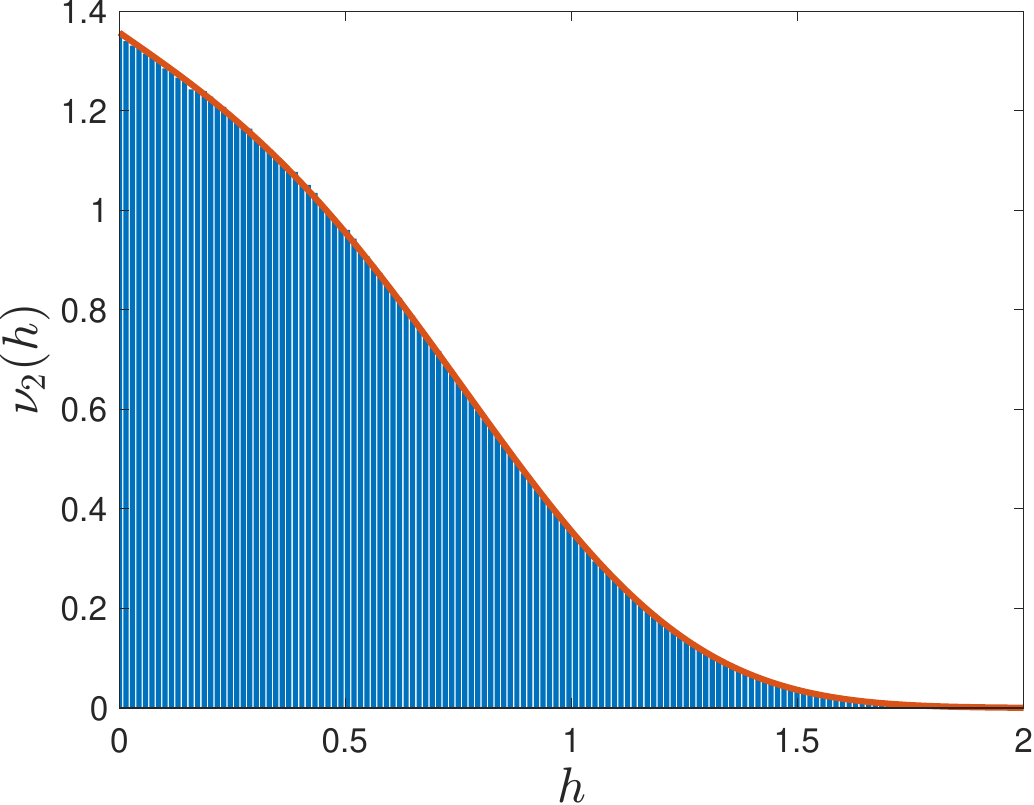}
  \caption{Scaling function for the number of site visits at the end-point of
    motion, $\nu_2(h)$ eq.~(\ref{eq:nu2}) curve in red, compared to simulation
    data, blue histogram.  Same simulation
    as Fig.~\ref{fig:nu1}}\label{fig:nu2}
\end{figure}

\ifheaders
\subsection*{Distributions functions of the model}
\fi
Remarkably, the equations~(\ref{eq:motion},~\ref{eq:density}) in one dimension
lead to an explicitly solvable model~\cite{Toth, DUMAZ20131454} involving time
scaling in $t^{2/3}$ or $t^{1/3}$ for distribution functions; the Langevin-like
eq.~(\ref{eq:motion}), does not lead to Brownian scaling in $t^{1/2}$. The exact
solutions displayed in~\cite{DUMAZ20131454} show that two important
distributions of physical variables exhibit scaling forms:
\begin{eqnarray}
  \rho_1(t,x) &=& t^{-2/3} \nu_1( x t^{-2/3}) \label{eq:scale1}\\
  \rho_2(t,h) &=& t^{-1/3} \nu_2 (h t^{-1/3} )\label{eq:scale2}
\end{eqnarray}
$\rho_1(t,x)$ is the distribution of displacement of the process after time $t$;
in the original polymer problem of~\cite{Amit} it is the distribution of
end-to-end separations. $\rho_2(t,h)$ is the distribution of $h(t)=L(t,X(t))$,
the density of previous visits to the endpoint of the polymer. The scaling
functions are
\begin{eqnarray}
  \nu_1(x) &= &\sum_{k=1}^{\infty}  \frac{p_k}{2} \delta'_k f_{2/3} (\delta'_k
                |x|) \label{eq:nu1}\\
  \nu_2 (h) &=  &\frac{2 \cdot 6^{1/3}\sqrt{\pi}}{\Gamma{(1/3)}^2} e^{-(8 h^3)/9}
                  U(1/6, 2/3, (8h^3)/9) \label{eq:nu2}
\end{eqnarray}
$f_{2/3}(x)$ is the Mittag-Leffler function. $\delta'_k$ and $p_k$ are
calculated from the k'th zero of the derivative of an Airy
function~\cite{DUMAZ20131454}, $U$ is a confluent hypergeometric function of the
second kind~\cite{NIST}. {The two functions
  eq.~(\ref{eq:nu1},~\ref{eq:nu2}), are rather distinctive visually and are far
  from Gaussian. They are plotted, in red, in Figs.~\ref{fig:nu1},~\ref{fig:nu2} }.

\ifheaders
\subsection*{Scaling argument for motion}
\fi
\ifanswers
A scaling argument~\cite{Pietronero}, allows one to find the exponents
describing the solutions.  Let motion occur on the length scale $t^{\alpha}$ in
time $t$. Then $L$ must be of the form
\begin{equation}
  L(t,x) \sim  t^{(1-\alpha)} g(x/t^{\alpha}) \label{eq:L}
\end{equation}
{with an unknown scaling function $g(x)$.}
The driving force from eq.~(\ref{eq:motion}), $-\nabla L(t,x)$, then scales as
$t^{(1-2 \alpha)} g'(x/t^{\alpha})$. If this force acts over the time $t$, it
generates motion $ \sim t^{(2-2\alpha)} $. It is natural that this motion is
comparable to the total extent of the motion, $t^{\alpha} $, so
$t^{(2-2\alpha)}$ = $t^{\alpha}$, and $\alpha=2/3$.
 \fi

\ifheaders
\section*{Event-Chain Monte Carlo}
\fi
Non-reversible, event-chain Monte Carlo methods are lifted variants of reversible Monte Carlo,
where a single ``active'' particle is mobile, and at (to be determined) event
times transfers motion to another particle, thereby avoiding the rejection step
of reversible Monte Carlo methods~\cite{Kapfer}. We here consider event-chain
algorithms suitable for simulation of models with continuous
potentials~\cite{Manon, Faulkner}.

The total energy function is broken into a sum of ``factor potentials''. Each of
these factors is then able to veto the motion of the active particle when a
stochastic criterion, is violated. {Vetos in the formulation of
  non-reversible Monte Carlo correspond to moments when a classical reversible
  algorithm would generate a rejection}. At the moment of veto, motion then transfers
to another particle {allowing forward motion to continue}. In the case of the
sampling of a chain, each particle is in direct interaction with its two
neighbours, the factor potentials are then just the contribution of each bond to
the total energy; on a veto event the activity jumps to one of the two neighbours.

Particularly exciting results for event-chain simulation were found for low-dimensional
XY spins~\cite{Lei_2018} where it was argued that in one dimension the dynamics
are characterized by a dynamic exponent, $z=1/2$. This exponent links the
relaxation time $\tau$ (in sweeps) of a system $N$ spins via $\tau\sim
N^{z}$. This result is smaller (hence, better) than is found in conventional
reversible Monte Carlo or molecular dynamics where one finds $z=2$ or
$z=1$. This exceptional scaling was generalized to more general models,
including hard spheres and Lennard-Jones chains where a detailed study of the
dynamics was performed~\cite{LeiFF}.  As noted in~\cite{LeiFF}, a dynamic
exponent of $z=1/2$, requires that the motion is indeed characterised by
hyperdiffusion, $x\sim t^{2/3}$. Rather remarkably, independent of the
underlying physical system the large-scale dynamic processes obey identical
hyper-diffusive dynamics.  We also note that scaling of the form
eq.~(\ref{eq:scale2}) was found for the {number of visits to the origin
  during a simulation} (see in
particular~\cite{LeiFF} Fig.~9).

Very similar phenomenology has recently been demonstrated in a lattice model, a
lifted version of the widely studied simple exclusion
process~\cite{DERRIDA199865, essler2023lifted}. {This variant of TASEP was designed to mimic certain
aspects of non-reversible Monte Carlo, in particular, the paper~\cite{essler2023lifted} introduces a new backward transfer of the active particle
inspired by the factor field construction from~\cite{LeiFF}}. Here too, strong numerical
evidence,  via the Bethe ansatz, points to a non-trivial scaling of
solutions involving $t^{2/3}$, {without giving access to scaling functions}. We conclude that this unusual scaling in time is
observed in a wide variety of physical models (with very different underlying
interactions) undergoing non-reversible dynamics subject to balance, without
detailed balance.

\ifheaders
\fi

%
\ifheaders
\subsection*{Self-repelling motion and non-reversible simulation}
\fi

We now simulate with an event-chain Monte Carlo an elastic chain at temperature, $T=1$, with harmonic
springs of unit strength, starting with a zero-temperature configuration. Our energy function is thus {\begin{equation}
  E= \frac{1}{2}\sum_i{(y_i -y_{i+1})}^2
  \end{equation}
}
{For concreteness, the particles move in only the $+y$ direction. It is this choice of a unidirectional update rule that
breaks detailed balance.} The
algorithm proposes, at each time step, a displacement of a mobile particle
which is vetoed after motion by a distance $O(1)$. \acm{We implement the
lifted event chain Monte Carlo algorithm as described by 
\cite{Faulkner}: At each moment a single particle is moving, its interaction
potential with each neighbour is calculated, 
until the randomly generated energy change of Ref.~\cite{Faulkner}, eq.~(45) is violated;   one of the neighbours is then selected as the target of a collision. 
After the collision, the newly chosen particle
starts to move and the process repeats.}

Consider now a section of the
chain, of $l+1$ sites of the system $[0,l]$, which has been uniformly stretched,
after running the algorithm for some time. For this to have occurred, there must
have been more visits to the site labelled $l$ than to the site $0$, for
instance for l=4 the zig-zag trajectory $(01234) (32) (1234) (3) (234) (34) $
generates a stretched chain with the number of visits for each site,
$L_i= [1, 2, 4, 6,\dots]$. This stretching then influences the future motion of
the active particle. Such a stretch leads to a higher veto rate for transfers of
the activity to the left of the chain, opposite to the gradient in
$L_i$. However, this is exactly the phenomenology of eq.~(\ref{eq:motion}). If
the exact details of the local update rule are not crucial in the dynamics and
some form of large-scale universality applies
eqs.~(\ref{eq:motion},~\ref{eq:density}) are clearly candidates for a
coarse-grained description of the motion generated by event-chain Monte
Carlo. In event-chain simulation in higher dimension, we understand that
relative motion of particles leads to a heterogeneous buildup in stress which
will then back-react on the motion of the active particle, leading to a very
similar coupling of motion and history, {though presumably the tensorial nature
  of stress in higher dimensions will change the exact mapping onto the
  self-avoiding walk}.

\ifheaders
\section*{Numerical results}
\fi
\ifheaders
\subsection*{Elastic chain}
 \fi

We generate data for the simulation of an elastic chain of $N=2^{26}$ particles linked
by harmonic springs in a periodic configuration. We initialise the chain and then perform $m=2^{27} \approx 1.34 \times 10^8$ simulations 
starting the mobile particle at  $i=0$. {In our comparisons we will treat the index variable,
  $i$,  of
  the chain as a continuum variable and will compare it to the chain motion $x$
  in the mathematical theory of the continuum stochastic process}.
  {Each simulation
starts with the chain in its ground state: $y_i =0$ for all $i$. This choice is
imposed by the fact that we wish to compare to the analytic calculations on the
self-avoid walk.}
All simulations are too short in time to be
influenced by the choice of a finite $N$, or to see wrap-around of trajectories.
{Again we note that the mathematical literature does not treat finite size effects, and we wish in this
  paper to make precise numerical comparisons with known analytic expressions.}
As is usual in event-driven Monte Carlo we simulate for an imposed chain length
which we take here as $t=10^6$. The mean-free path is slightly greater than unity, so
this corresponds to very nearly $10^6$ collisions {for each of the $m$ simulations}.

During each simulation, we cumulate
the number of visits to each site by the active particle, {in a manner analogous to  eq~(\ref{eq:disc}) } 
which we again denote $L_i(t)$. At the end of
each of the $m$ simulations we save the index of the final active particle (which in the continuum
limit corresponds to a displacement $x$), as
well as the number of visits of the active particle to the final chain position
{corresponding to the variable
$h$ of eq.~(\ref{eq:nu2})}. After the simulation, the data is binned generating empirical
distributions, that we compare with eqs.~(\ref{eq:nu1},~\ref{eq:nu2}), see
Figs.~\ref{fig:nu1},~\ref{fig:nu2}. {The red
curves are the analytic results; blue shading corresponds to binned data from the Monte
Carlo data. Fig.~\ref{fig:nu1} corresponds to some 40,000 bins in the data,
which give the appearance of a continuous curve.}
The
curves represent probability distributions and are thus automatically normalized
to unit area.  The only free variable in comparing the analytic expressions to
the binned numerical data is the choice of the horizontal scale.  We fix
this one scale by imposing that {$\langle |x| \rangle $} of the empirical
curve matches the theoretical expression. {Fig.~\ref{fig:nu1} is centered at
zero due to the pressure theorem of~\cite{Manon} which links the average motion
to the stress in the physical system.}

We
find excellent agreement between theory and numerics. We note, in particular,
that as predicted by theory the distribution of Fig.~\ref{fig:nu1} has a singularity at
the origin.
The empirical curve, Fig.~\ref{fig:nu2}, for 
$\rho_2$ at $h=0$ is systematically lower than the red, theoretical curve, this
corresponds to simulations that finish on an as-yet unvisited site. This empirical curve,
thus converges weakly to the theory, with the bar at $h=0$ squeezed to zero
width at the origin 
as $t$ increases.
We also performed simulations starting from a pre-equilibrated
physical system. In this case, {we find an identical distribution} to 
that plotted in Fig.~\ref{fig:nu1},  with however, a change of scale in
the axes, see also~\cite{DUMAZ20131454}.

\ifheaders
\subsection*{Lifted TASEP}
\fi

Finally, we perform a numerical study of the lifted TASEP model~\cite{essler2023lifted}. We consider a system of $N$ particles on a periodic
lattice of length $2N$. {This new variant on the TASEP is a model of
  impenetrable particle motion on a lattice, in which collisions cause transfers
  of velocity. The modification in the calculation of~\cite{essler2023lifted} is
  the addition of a backward transfer of particle motion. For a special value of
the backward transfer rate the lifted TASEP displays accelerated relaxation dynamics in
density-density correlations in a manner which is similar to that displayed in
continuum particle models\cite{LeiFF}. We study this TASEP model only at this
specific value of the backward rate where the dynamics is highly accelerated.}

\begin{figure}
  \includegraphics[width=.65\columnwidth]{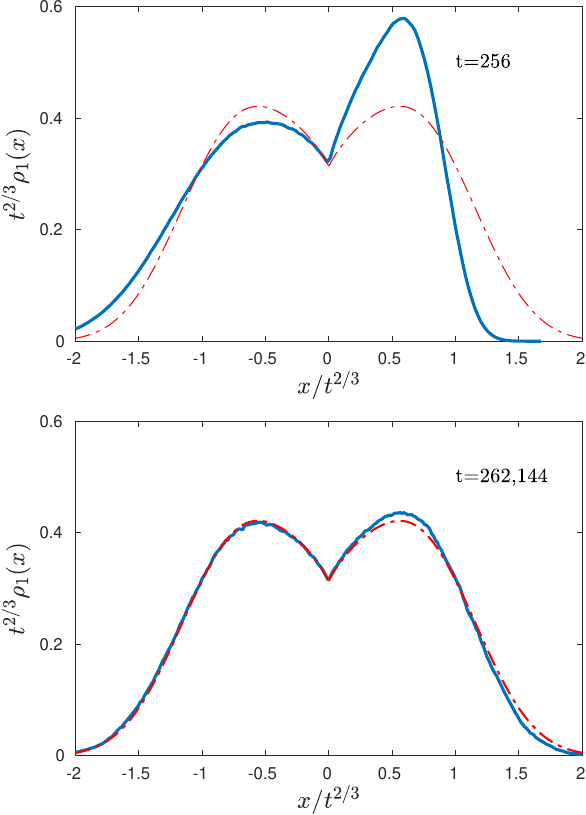}
  \caption{Distribution of displacements of the activity $\rho_1(t,x)$, for lifted
    TASEP\@. Top, a short simulation of $t=256$ steps generates a skewed
    distribution. Bottom, a longer time scale of $t=262,144$ generates a
    distribution which is close to that found in Fig.~\ref{fig:nu1}, while still
    displaying a very slight left-right asymmetry. Scaled curve of $\nu_1(x) $
    red/dashed.}\label{fig:tasep}
\end{figure}

\begin{figure}
  \includegraphics[width=.65\columnwidth]{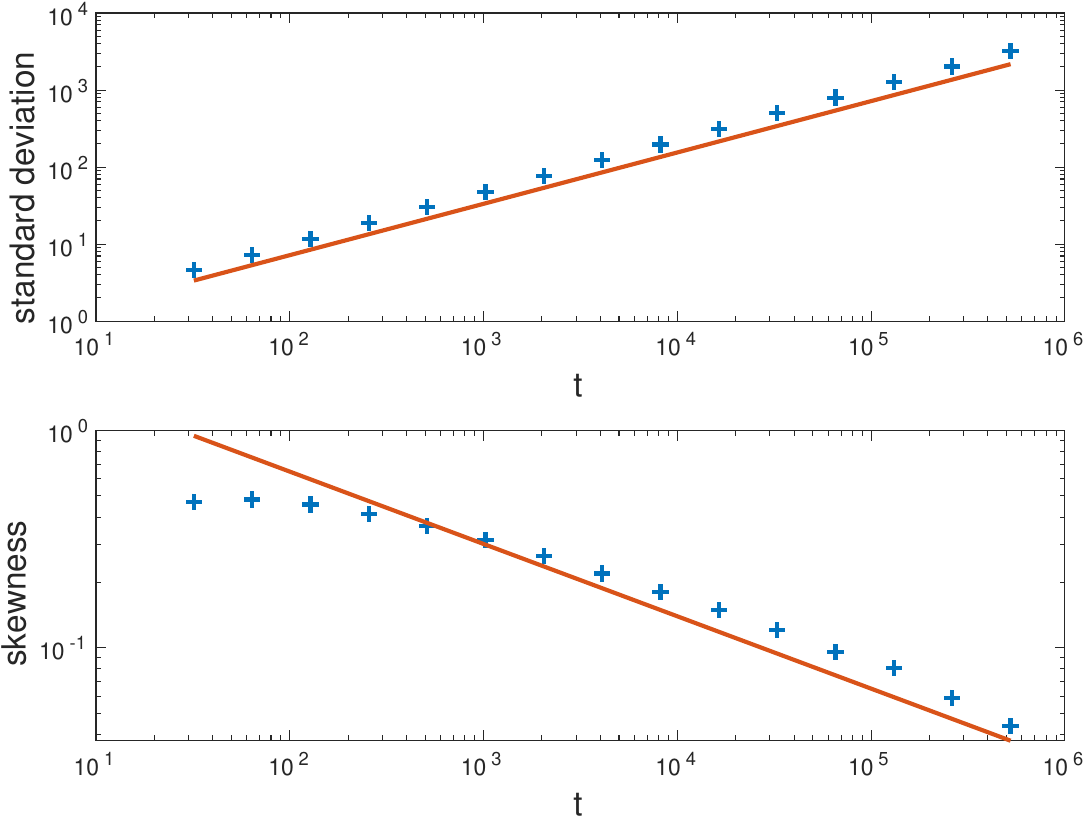}
  \caption{Evolution of moments of $\rho_1(t,x)$ for lifted TASEP\@. Top: standard
    deviation of the distribution, compared to evolution in $t^{2/3}$. Bottom,
    skewness (normalized third moment) compared to $t^{-1/3}$. The decrease in
    skewness corresponds to a distribution which progressively becomes more
    symmetric.}\label{fig:moments}
\end{figure}

\begin{figure}
  \includegraphics[width=.65\columnwidth]{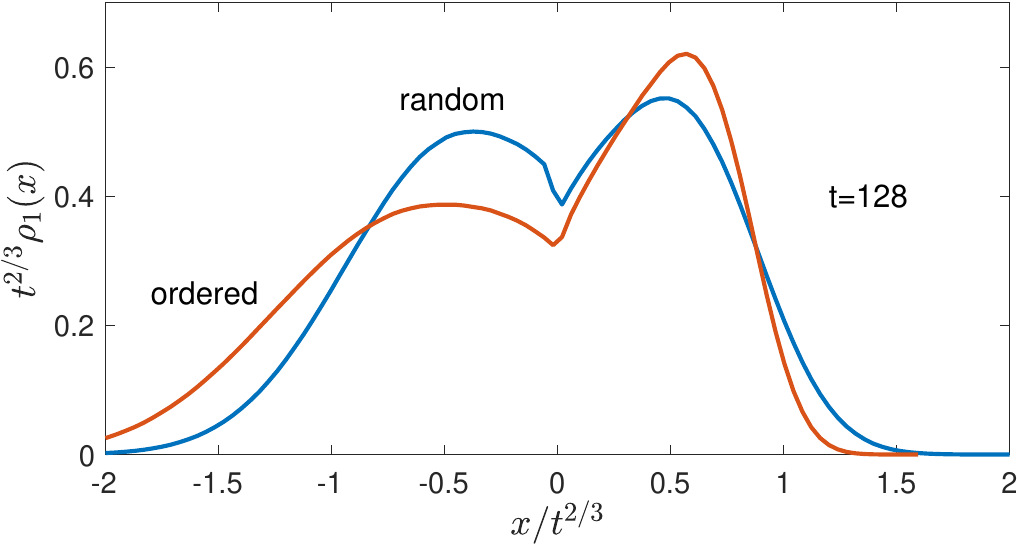}
  \caption{{Distribution of displacements of the activity $\rho_1(t,x)$, for lifted TASEP\@. We
      either start the simulation either in an ordered state, or alternatively start with a
      random state drawn from the equilibrium distribution. The randomized starting condition displays
      a distribution with smaller skewness. Data not rescaled to fit theoretical
    curve.}}\label{fig:doubletasep}
\end{figure}

We initialise the system in a crystal of equally spaced particles. \acm{Again,
  we will treat the index variable, $i$, of the particles of the chain as a continuum variable
  and will compare it to the end motion of a  chain $x$.}  Simulations on short times give
rise to an asymmetric distribution from for $\rho_1(x)$ (Fig.~\ref{fig:tasep},
Top), but longer simulations give a very slow convergence to a more symmetric
form (Fig.~\ref{fig:tasep}, bottom). We conclude that the lifted TASEP displays
very similar phenomenology to the harmonic chain, and is also in the same
dynamic universality class as true self-avoiding motion. We also confirmed, for
lifted TASEP, the expected scaling of displacement of the activity in $t^{2/3}$,
Fig.~\ref{fig:moments}~(top).  Fig.~\ref{fig:moments},~(bottom) shows that the
distribution of displacements is strongly skewed for short times; it is only
after a long simulation that the back-forwards symmetry is established When we
{average over} configurations which are generated {from the equilibrium
  distribution}, the initial asymmetry is weaker, {Fig.~\ref{fig:doubletasep}}
though not
zero. 

\ifheaders
\section*{Conclusions}
\fi
To conclude, we have presented {numerical evidence} that the evolution of a system
subject to non-reversible Monte Carlo is directly linked to the continuum limit
of a growth model~\cite{Amit}. We performed extensive simulations using an event
chain algorithm and compared the resulting distributions to those calculated
in~\cite{Toth,DUMAZ20131454,tothProc}. We find excellent agreement, for both a harmonic
chain, and for the non-harmonic lifted TASEP, and so conclude that the
non-reversible algorithm is indeed a realisation of a true self-repelling
motion. {We have shown that several {\sl very different} physical systems are
  members of the same dynamic universality class. It would be of great
  interest to understand why this is the case, and whether other 
  systems display the same scaling behavior. As noted by~\cite{DUMAZ20131454} this
  universality class is distinct from KPZ (Kardar-Parisi-Zhang), even though
   similar special functions appear in the solutions}. \acm{On the algorithmic
   side we note that ``worm'' algorithms \cite{worm} also seem to have
   similarities to growing polymers, and have also led to
   substantial speed-ups in the simulation of classical and quantum spin models.}

Event-chain methods, including factor fields, have been generalized to higher
dimensions~\cite{Maggs_2022}. It would be of interest to transfer the formalism
of the present letter to such systems, perhaps using the approach of~\cite{toth-net}. Applications in
Monte Carlo simulation also require generalisations of the continuum
equations to include extra, drift, terms due to coupling to external stresses~\cite{Manon}.


The code used to simulate the two physical systems, as well as the plotting and analysis code
is available from {\url{https://github.com/acmaggs/Avoid}}.
%
\ack
{The author would like to thank Werner Krauth for extensive discussions on
  the subject of non-reversible simulation and lattice models.} 

\bibliographystyle{iopart-num} 
\ifanswers
\section*{References} 
\bibliography{avoid}
\end{document}
